\documentstyle[epsf,aps]{revtex}

\begin{document}

\wideabs{

\title{
Spin-Peierls transition of the first order
in $S=1$ antiferromagnetic Heisenberg chains
}

\draft

\author{Hiroaki Onishi}
\address{
Department of Earth and Space Science,
Graduate School of Science, Osaka University,\\
Machikaneyama-cho 1-1, Toyonaka, Osaka 560-0043, Japan
}
\author{Seiji Miyashita}
\address{
Department of Applied Physics,
Graduate School of Engineering, University of Tokyo,\\
Hongo 7-3-1, Bunkyo-ku, Tokyo 113-8656, Japan
}

\date{1 Sep 2001}

\maketitle

\begin{abstract}
We investigate a one-dimensional $S=1$ antiferromagnetic Heisenberg model
coupled to a lattice distortion by a quantum Monte Carlo method.
Investigating the ground-state energy of the static bond-alternating chain,
we find that 
the instability to a dimerized chain
depends on the value of the spin-phonon coupling,
unlike the case of $S=\frac{1}{2}$.
The spin state is the dimer state or the uniform Haldane state
depending on whether the lattice distorts or not, respectively.
At an intermediate value of the spin-phonon coupling,
we find the first-order transition between the two states.
We also find the coexistence of the two states.
\end{abstract}


}

\section{INTRODUCTION}
\label{sec:introduction}

A spin-Peierls system has been one of the most fascinating topics
of the low-dimensional quantum spin system.
The one-dimensional $S=\frac{1}{2}$ antiferromagnetic Heisenberg chain
shows a spontaneous lattice dimerization
due to the spin-phonon coupling at low temperature,
where the strength of the antiferromagnetic coupling alternates
between strong and weak values.
For $S=\frac{1}{2}$,
there have been numerous studies
on the ground state of the spin-Peierls system.
It has been shown
by a bosonization method~\cite{Cross-Fisher,Nakano-Fukuyama}
that
the bond alternation $J_i=J[1+(-1)^i\delta]$
causes the magnetic energy  gain proportional to $\delta^{4/3}$
for small $\delta$.
Therefore,
the magnetic energy gain always exceeds the elastic energy for small $\delta$,
because the elastic energy is proportional to $\delta^2$.
Thus,
the dimerized state is always realized in the ground state
regardless of the value of the spin-phonon coupling.
This mechanism has been confirmed in numerical works~\cite{MC-self,DMRG-self},
where the dimerized lattice configuration is obtained
in the balance of the magnetic energy and the elastic energy
in the ground state.
Properties at finite temperature have been also
investigated~\cite{ph-1,ph-2,spqmc},
where the distribution of the lattice distortion
in the thermal equilibrium state is obtained.
Recently,
impurity-induced long range orders have been found
and their physical origin has been discussed extensively~\cite{imp_review}.

On the other hand,
only a small amount of study on the spin-Peierls transition for
$S>\frac{1}{2}$ cases has been done.
In the case of integer spin antiferromagnetic Heisenberg uniform chains,
as pointed out by Haldane~\cite{Haldane},
an energy gap exists between the ground state and the first excited state,
and the spin-spin correlation function decays exponentially
contrary to the case of half-odd-integer spin chains.
Because of the qualitative difference of the ground-state property
between half-odd-integer spin and integer spin,
the nature of the spin-Peierls transition in integer spin systems
is expected to be different from that in half-odd-integer spin systems.
Guo {\it et al.} investigated the magnetic energy gain due to the dimerizaion
for $S=\frac{1}{2},1,\frac{3}{2}$ and $2$
by an exact diagonalization method~\cite{spgsdiag}.
They pointed out that
the transition in systems with integer spin occurs
only for large value of the spin-phonon coupling
while
the transition occurs
regardless of the value of the spin-phonon coupling
in systems with half-odd-integer spin.
There this difference was discussed
from a view point of the valence-bond-solid (VBS) picture.
However,
the nature of the phase transition was not discussed in detail.

\begin{figure}
\centering \epsfxsize=60mm \epsfbox{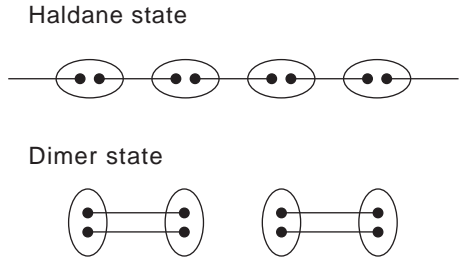}
\vspace*{5mm}
\caption{
The conceptual configurations of the VBS state for $S=1$.
}
\label{fig:VBS}
\end{figure}

In the spin-$S$ antiferromagnetic Heisenberg bond-alternating chains,
successive quantum phase transitions occur at $2S$ gapless points
in the range $-1\leq\delta\leq1$%
~\cite{Affleck-Haldane,S1BA_KaTa,S1BA_YaOsMi,S1BA_Ya,S1BA_KoTa}.
The transitions can be considered as ones
between the different VBS states,
as shown in Fig.~\ref{fig:VBS}.
In the case of $S=1$,
the continuous transition between the Haldane state and the dimer state occurs
at the critical point $\delta_c \simeq 0.26$~\cite{S1BA_KaTa}.

In this paper
we study the case of $S=1$
by a quantum Monte Carlo method.
First
we consider the instability to a dimerized chain in the ground state.
In the static bond-alternating chain,
the ground state depends on
the amplitude of the bond alternation
as we mentioned above.
Therefore,
there are two possibilities for the instability to the dimerized chain,
namely,
the chain is unstable toward the dimerized chain with a small bond alternation
where the spin state is the Haldane state,
or toward the dimerized chain with a large bond alternation
where the spin state is the dimer state.
To which state the system changes
depends on the value of the spin-phonon coupling.
We also find that
the uniform Haldane state and the dimerized state coexist
in the ground state at an intermediate value of the spin-phonon coupling.
On the other hand,
we find that the dimerized state is more stable than the uniform Haldane state
at finite temperature,
which can be attributed to an entropy effect.
Furthermore, we investigate an open chain with fixed boundary bonds,
by which we manipulate the configuration of the bonds at edges.
Locating bond alternation at one end
and uniform configuration at the other end,
we force the system to have the two regions,
and observe the situation where
the uniform Haldane state and the dimerized state coexist in the chain.
We investigate the domain wall between the two regions.

This paper is organized as follows.
In Sec.~\ref{sec:model and method}
we explain a model and a numerical method.
We also refer to the order parameters.
In Sec.~\ref{sec:dimerized ground state}
the magnetic energy gain due to the dimerization is investigated
in order to study the instability to a dimerized bond structure
in the ground state.
In Sec.~\ref{sec:thermal fluctuation}
the distribution of the bond distortion at finite temperature and
the effect of the entropy on the coexistence of the two states
are investigated.
In Sec.~\ref{sec:coexistent state}
the domain-wall configurations
between the uniform Haldane state and the dimerized state are investigated.
We summarize our results
in Sec.~\ref{sec:summary}.
%
%

\section{MODEL AND METHOD}
\label{sec:model and method}

The spin-Peierls system is a spin-phonon coupled system described by
\begin{equation}
H = J\sum_{i=1}^N
    (1+\alpha\tilde{\Delta}_i) {\bf S}_i\cdot{\bf S}_{i+1}
  + {\tilde{k}\over2}\sum_{i=1}^N \tilde{\Delta}_i^2,
\end{equation}
%
%
where the distortion of the exchange coupling is
assumed to be proportional to the lattice distortion,
that is, $\Delta J_i/J = \alpha\tilde{\Delta}_i$.
Here $\alpha$ is a coefficient
that gives the distortion of the exchange coupling
and is negative $(\alpha < 0)$
because the shrink of the lattice causes the increase in the exchange coupling.
In this paper,
for the simplicity of notation,
we use the change of the exchange coupling
$\Delta_{i} \equiv \alpha\tilde{\Delta}_i$
instead of $\tilde{\Delta}_i$ as a variable for the distortion
and also scale the elastic constant as $k \equiv \tilde{k}/\alpha^2$,
which leads to
\begin{equation}
H = J\sum_{i=1}^N
    (1+\Delta_i){\bf S}_i\cdot{\bf S}_{i+1}
  + {k\over2}\sum_{i=1}^N \Delta_i^2.
\label{eq:hamsp}
\end{equation}
We study this Hamiltonian~(\ref{eq:hamsp}) for $S=1$
by a world line quantum Monte Carlo (QMC) method
in order to investigate thermodynamic properties of the model.
We use a quantum Monte Carlo loop algorithm
with continuous imaginary time~\cite{loop-1,loop-2,loop-c}.
This algorithm is effective for studying the property at very low temperature.
We have studied
the $S=\frac{1}{2}$ system with the bond distortion
by the loop algorithm,
where we investigated how the dimerization develops
as a function of the temperature
and also the effect of the bond fluctuation at high temperature
where the averaged bond structure is still uniform~\cite{spqmc}.
The system is updated in the following way.
Starting from an arbitrary spin and bond configuration,
the spin configuration is updated by the loop algorithm
with the fixed bond configuration.
The loop algorithm of a general spin is constructed
as the extension of the $S=\frac{1}{2}$ format~\cite{loop-s}.
The bond configuration $\{\Delta_i\}$ is updated
by the Metropolis algorithm
with the fixed spin configuration.
Here we treat the bond distortion as a classical quantity,
although the quantum phonon effect is also
an interesting problem~\cite{ph-1,ph-2}.
These two updated procedures are carried out alternately,
therefore
we obtain the thermal equilibrium state both of spin and bond.
In the present study,
we impose the following constraint in the simulation,
\begin{equation}
\sum_{i=1}^N \Delta_i = 0,
\label{eq:length}
\end{equation}
in order to fix the total length of the lattice.
Besides, we restrict the bond distortion to $-1<\Delta_i<1$
in order to confine the exchange coupling to be antiferromagnetic.
The periodic boundary condition is adopted except in
Sec.~\ref{sec:coexistent state},
where we adopt the open boundary condition with fixed boundary bonds.
We set the Boltzmann factor $k_{B}=1$ 
and use it as the unit of the energy.
We fix the uniform exchange coupling $J=1$
and investigate cases with various values of $k$.

We investigate the ground state energy of static bond-alternating chains
in order to determine how much energy the system gains
from the magnetic interaction in a given bond configuration.
In order to obtain the energy,
we perform a QMC method without a bond update.
The initial $10^3$ Monte Carlo steps (MCS) are discarded
to obtain the thermal equilibrium state
and the data of physical quantities are sampled in the following $10^4$ MCS.
For the study of bond-fluctuating chains,
larger simulations such as $10^6$ MCS are performed to obtain
good convergence of the data.
The sampled data is divided into $10$ bins,
and the error bar is estimated from the standard deviation in this set of data.

\subsection{The order parameters}
\label{sec:the order parameters}

In this paper
we study orderings of the bond configuration and the spin configuration.
In order to investigate the bond ordering,
we introduce the bond staggered order parameter
\begin{equation}
\Delta_{\rm sg}^2 = \left({1 \over N}\sum_{i=1}^N (-1)^i \Delta_i\right)^2,
\label{eq:bsg2}
\end{equation}
which represents how the bond alternates in the whole chain.
When the bond configuration takes an alternating arrangement on average,
the spin state would change
according to the amplitude of the bond alternation.
In the static bond-alternating chain,
the hidden $Z_2 \times Z_2$ order exists
in the Haldane state, while not in the dimer state.
The hidden order is detected by the string order parameter
introduced by den Nijs and Rommelse~\cite{Nijs-Rommelse},
\begin{equation}
O_{\rm string} = \lim_{|i-j|\rightarrow\infty} C_{\rm string} (i,j),
\label{eq:str-original}
\end{equation}
where
\begin{equation}
C_{\rm string} (i,j) =
S_i^z \exp \left[ i \pi \sum_{k=i}^{j-1} S_k^z \right] S_j^z.
\label{eq:strcor}
\end{equation}
In order to investigate whether the spin state takes
the Haldane state or the dimer state,
we observe the string correlation by the following quantity~\cite{S1_YaMi},
\begin{equation}
O_{\rm LR} =
\frac{1}{N_{\rm s}}
\sum_{i=1}^N \exp \left[ i \pi \sum_{k=1}^i S_k^z \right] S_i^z,
\label{eq:str}
\end{equation}
where $N_{s}$ is the number of nonzero spins,
and $N_{s}=\sum_{i=1}^N |S_i^z|$.
Then we calculate $\langle O_{\rm LR}^2 \rangle$:
\begin{equation}
\langle O_{\rm LR}^2 \rangle
=
\frac{1}{N_{s}^2}
\sum_{i=1}^N \sum_{j=1}^N \langle C_{\rm string} (i,j) \rangle
\propto
\frac{1}{N_{s}}
\xi_{\rm string},
\label{eq:str2}
\end{equation}
which gives an estimate of the string correlation length $\xi_{\rm string}$.

\section{DIMERIZED GROUND STATE}
\label{sec:dimerized ground state}

\begin{figure}
\centering \epsfxsize=84mm \epsfbox{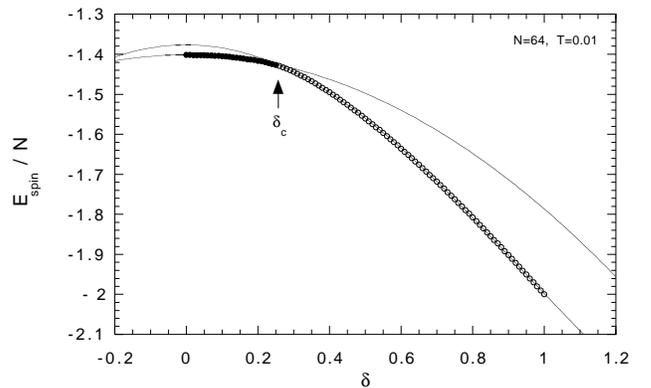}
\vspace*{1mm}
\caption{
The $\delta$ dependence of the ground-state magnetic energy
with the dimerization $\Delta_i = (-1)^i\delta$.
As a guide to the eye,
the values for $\delta<\delta_c$ and for $\delta>\delta_c$
are plotted with the solid circles and the open circles, respectively.
The solid lines denote fitting curves
for each region of $\delta$ on both sides of $\delta_c = 0.26$
as mentioned in the text.
}
\label{fig:e-d}
\end{figure}

\begin{figure}
\centering \epsfxsize=84mm \epsfbox{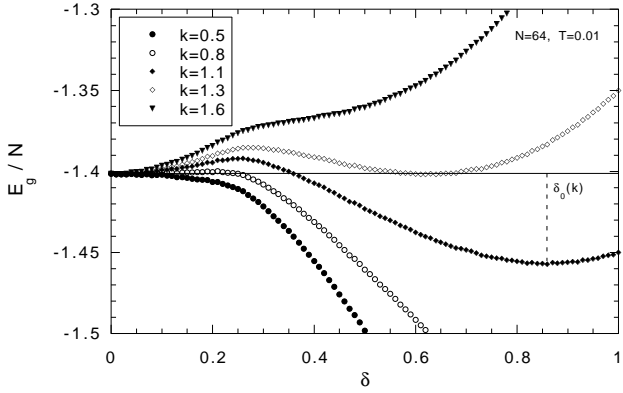}
\vspace*{1mm}
\caption{
The $\delta$ dependence of the ground-state energy for various values of $k$.
}
\label{fig:e-d-vk}
\end{figure}

In this section
we investigate the dependence of the ground-state energy
on the amplitude of the bond distortion by the QMC method.
This dependence gives an effective (or adiabatic) potential
for the bond distortion,
by which we can study the instability to the dimerized chain
in the ground state.
Here we consider static dimerized chains with an amplitude $\delta$,
\begin{equation}
\Delta_i = (-1)^i \delta.
\end{equation}
In Fig.~\ref{fig:e-d},
we show the magnetic energy as a function of $\delta$,
$E_{\rm spin}(\delta)$
for the system of $N=64$ and $T=0.01$.
By studying the size and temperature dependencies,
we confirm that
this system size is sufficiently large enough to obtain the ground-state energy
without the finite-size effect,
and this temperaure is sufficiently low enough to obtain the ground-state energy
without the finite-temperature effect.
The magnetic energy decreases monotonously as $\delta$ increases.
Besides,
we find a singular behavior in $E_{\rm spin}(\delta)$
at the critical point $\delta_c = 0.26$
where the phase transition occurs
between the Haldane state and the dimer state~\cite{S1BA_KaTa}.
The magnetic energy curve can be fitted well by an even function
in each region of $\delta$,
\begin{eqnarray}
E_{\rm spin}(\delta)/N =
\left\{
\begin{array}{l}
a  + b  \delta^2               \quad (\delta<\delta_c)  \\
a' + b' \delta^2 + c' \delta^4 \quad (\delta>\delta_c), \\
\end{array}
\right.
\label{eq:e-d_spin}
\end{eqnarray}
with
$(a, \, b) = (-1.40, \, -0.385)$
and
$(a', \, b', \, c') = (-1.38, \, -0.770, \, 0.148)$.
Adding the elastic energy,
$E_{\rm lattice}(\delta)/N=\frac{1}{2}k\delta^2$,
we obtain the ground-state energy as a function of $\delta$,
\begin{equation}
E_{\rm g}(\delta) = E_{\rm spin}(\delta) + E_{\rm lattice}(\delta),
\end{equation}
which is interpreted as an effective potential for the bond distortion.
The results are shown for some values of $k$ in Fig.~\ref{fig:e-d-vk}.
First let us consider the case of small $\delta$.
When $\frac{1}{2}k > |b|$,
the uniform configuration $\delta = 0$ gives a stable point at least locally.
On the other hand,
when $\frac{1}{2}k < |b|$,
the uniform configuration becomes unstable.
Here the instability to the dimerized chain occurs at $\frac{1}{2}k = |b|$.
However,
if we look at the shape of $E_{\rm g}(\delta)$ globally,
we find that
a minimum point of the potential jumps from $\delta = 0$
to $\delta_0 \, (>\delta_c)$ discontinuously.
Namely,
the energy shows the characteristic behavior of the first-order transition.
The fourth term $c'\delta^4$ suppresses the divergence
and gives a finite value $\delta_0$, which minimizes the total energy.
In the present study, however,
we restrict the value of $\delta$ to $|\delta| < 1$.
Therefore,
$\delta = 1$ gives the minimum value of the total energy
when $\delta_0 > 1$.
This constraint is introduced
in order to confine the exchange coupling to be antiferromagnetic.
Physically,
this constraint corresponds to a nonlinear increase
of the energy of the lattice distortion
or a nonlinear relation
between the lattice distortion and the change of the exchange coupling.
%
%

%
%
%
%
\begin{figure}
\centering \epsfxsize=84mm \epsfbox{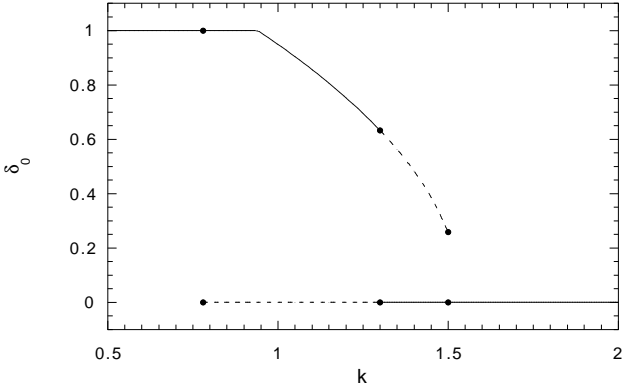}
\vspace*{1mm}
\caption{
The $k$ dependence of $\delta_0$, which minimizes the ground-state energy.
The solid line denotes a stable value as a function of $k$.
The dotted line denotes an unstable value as a function of $k$.
Here we restrict $\delta$ to within $\delta < 1$.
}
\label{fig:d0-k}
\end{figure}

The ground-state value $\delta_0$ as a function of $k$
is plotted in Fig.~\ref{fig:d0-k}.
For large values of $k$,
the energy increases monotonously as $\delta$ increases
and the ground state is the uniform Haldane state.
For intermediate values of $k$,
$E_{\rm g}(\delta)$ has two minimum points. 
The minimum $\delta=0$ corresponds to a uniform chain
where the spin state is the Haldane state,
while $\delta=\delta_0 \, (>\delta_c)$ corresponds to a bond-alternating chain
where the spin state is the dimer state.
In particular
the energies of the two minima are the same at $k=k_c=1.30$.
Here two spin-lattice configurations coexist in the ground state.
For small values of $k$,
the energy decreases monotonously
and the ground state is the dimerized state.
It should be noted that
$\delta_0$ is always larger than $\delta_c$.
Therefore,
we conclude that the spin state is the dimer state
whenever the latice is spontaneously distorted.
Namely,
in the ground state of the present system,
the Haldane state appears only in a uniform chain
and does not appear in a weakly bond-alternating chain.

In the coexistent range,
the maximum of the energy between the two minima
corresponds to an energy barrier $\Delta E$
between the uniform Haldane state and the dimerized state.
In particular around $k=1.3$,
the energy barrier is rather small,
i.e., $\Delta E \simeq 0.02$.
Therefore,
a large fluctuation is expected
at finite temperature higher than the height of this energy barrier,
i.e., $T \simeq 0.02$.

\section{THERMAL FLUCTUATION}
\label{sec:thermal fluctuation}

In this section
we study the bond distortion $\{\Delta_i\}$
at finite temperature by the QMC method.
First
let us study the distribution of the bond distortion.
At $T=0$,
the distribution $P(\Delta)$ is the delta function
with one peak at $\Delta=0$ for the uniform chain
or two peaks at $\Delta=\pm\delta_0$ with the same amplitude for the bond-alternating chain.
The peaks are broadened at finite temperaure
due to the thermal fluctuation.
At high temperature,
we find a broad single peak near at $\Delta = 0$.
On the other hand,
at low temperature,
we find a sharp distribution with one peak at $\Delta = 0$
or two peaks at $\Delta = \pm\delta_0$
according to the value of $k$.
Thus we confirm that
the dimerized ground state
obtained in the previous section is actually realized,
and it ensures that the lattice takes the dimerized configuration
although many other configurations are possible in principle.

\begin{figure}
\centering \epsfxsize=84mm \epsfbox{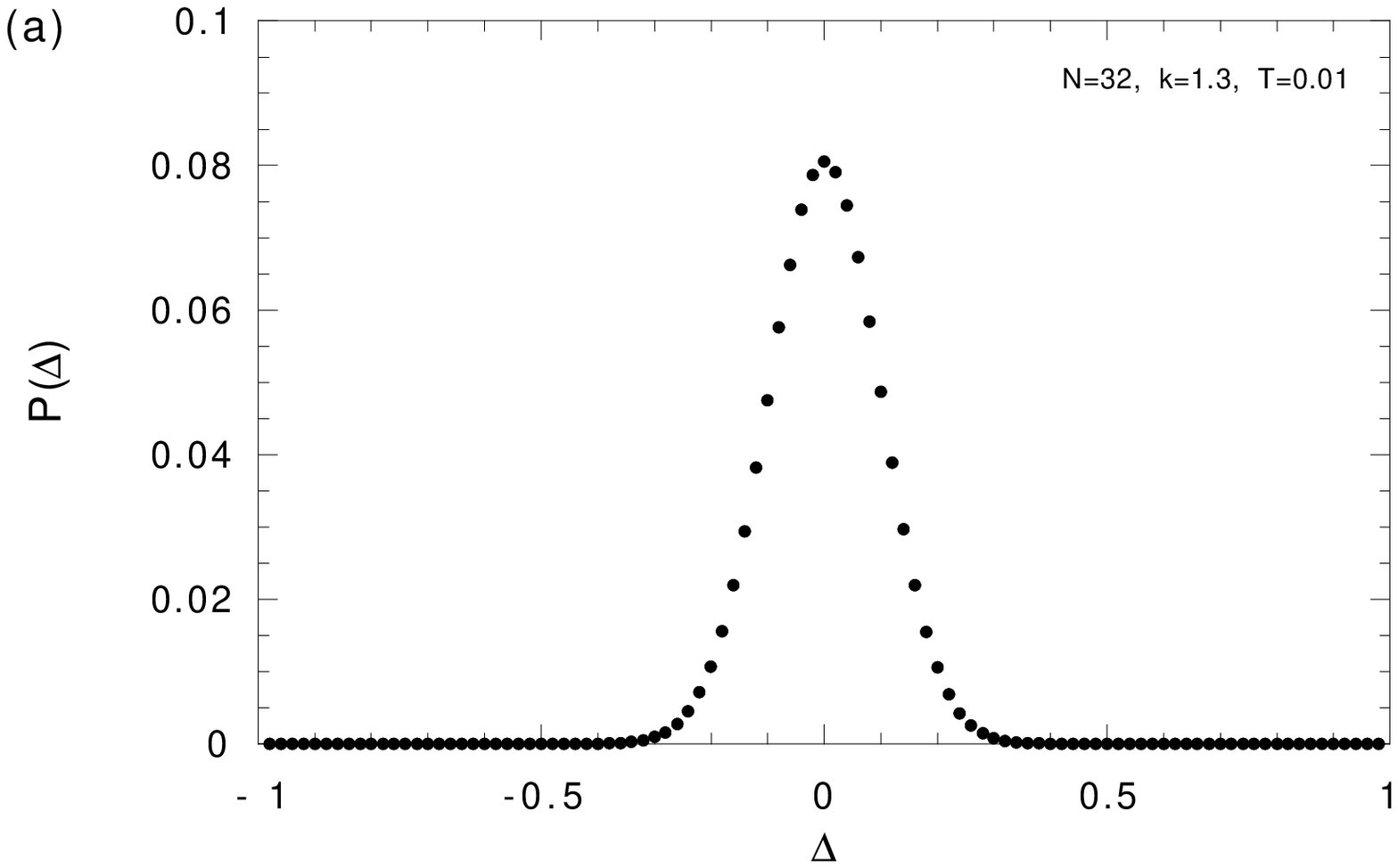}
\epsfxsize=84mm \epsfbox{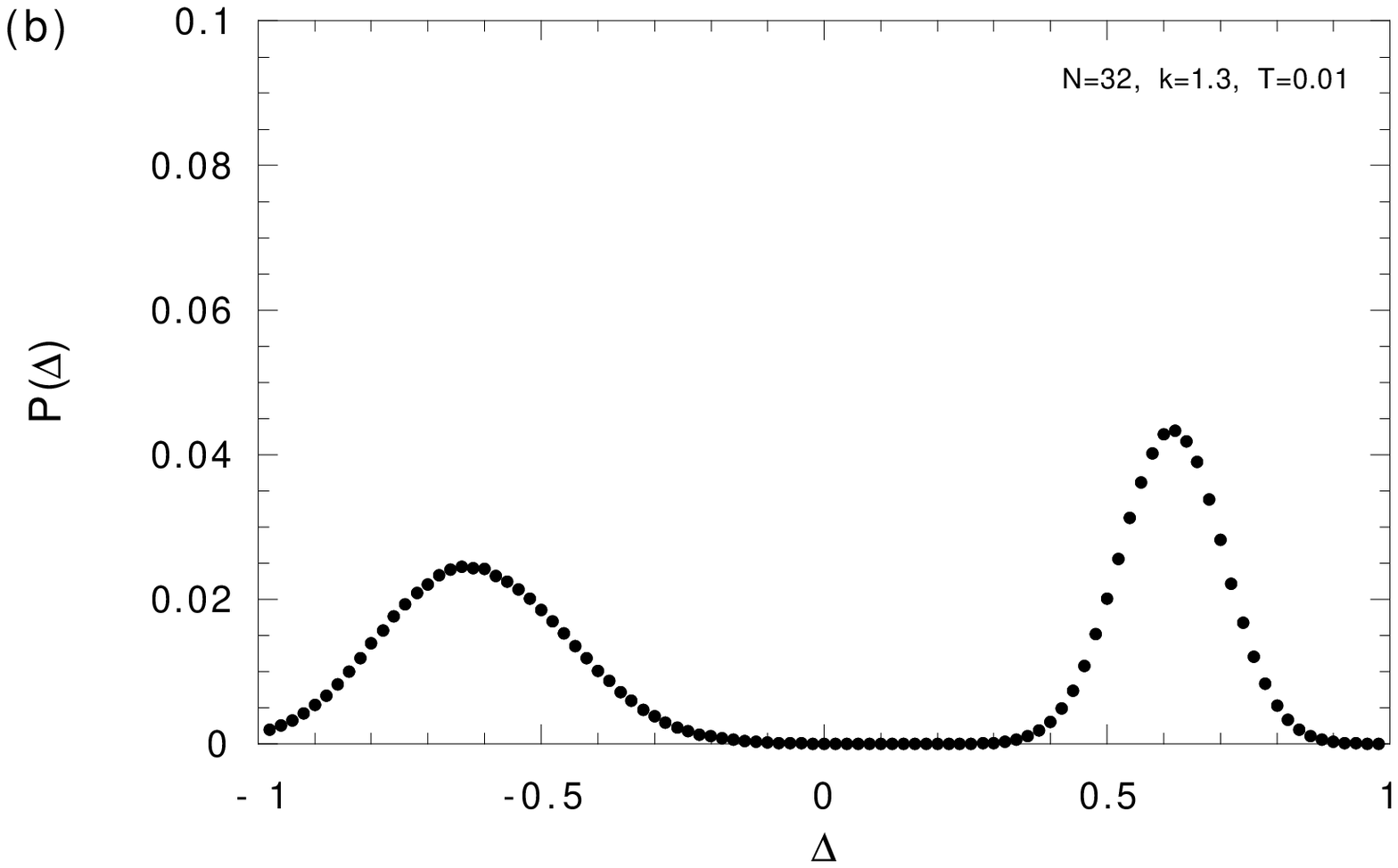}
\vspace*{1mm}
\caption{
The distribution function of the bond distortion
at $T=0.01$ for $k=1.3$ and $N=32$.
(a) corresponds to the uniform Haldane state
and (b) corresponds to the dimerized state.
}
\label{fig:p-d}
\end{figure}

As mentioned in the previous section,
the uniform Haldane state and the dimerized state coexist
in the ground state at $k = k_c$.
It is expected that the distribution is given
by a combination of each distribution.
However, the system takes either state at a time in the simulation.
In Fig.~\ref{fig:p-d},
we show the distribution function
for $k=1.3$ at a very low temperature $T=0.01$.
We find that
the system is trapped either in the uniform state [Fig.~\ref{fig:p-d}(a)]
or in the dimerized state [Fig.~\ref{fig:p-d}(b)].
Once the system is trapped in one state,
it is rather difficult to transfer to the other state
at the temperature lower than the energy barrier.
In Fig.~\ref{fig:p-d}(b),
we find an asymmetrical distribution for the dimerized state.
The amplitude at the positive $\Delta$ peak is larger
than that at the negative $\Delta$ peak.
This difference can be attributed to the energy difference of the state.
We have observed this asymmetry
in the case of $S=\frac{1}{2}$ as well~\cite{spqmc}.

\begin{figure}
\centering \epsfxsize=84mm \epsfbox{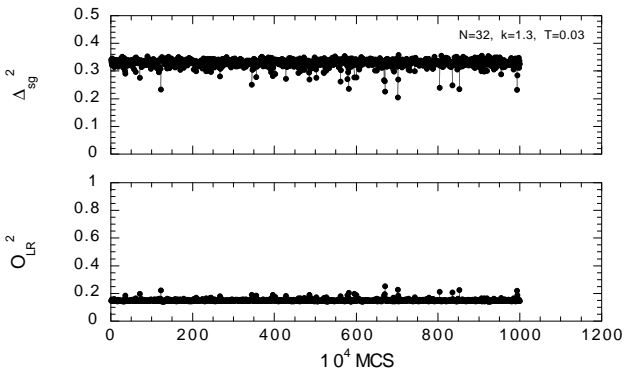}
\vspace*{1mm}
\caption{
The MCS dependence of the bond staggered order $\Delta_{\rm sg}^2$
and the string order $O_{\rm LR}^2$ at $T=0.03$ for $k=1.3$ and $N=32$.
The whole simulation is divided in blocks of $10^4$ MCS
and the thermal averaged value is calculated in each bin.
}
\label{fig:bsg2str2_k1.3}
\end{figure}

\begin{figure}
\centering \epsfxsize=84mm \epsfbox{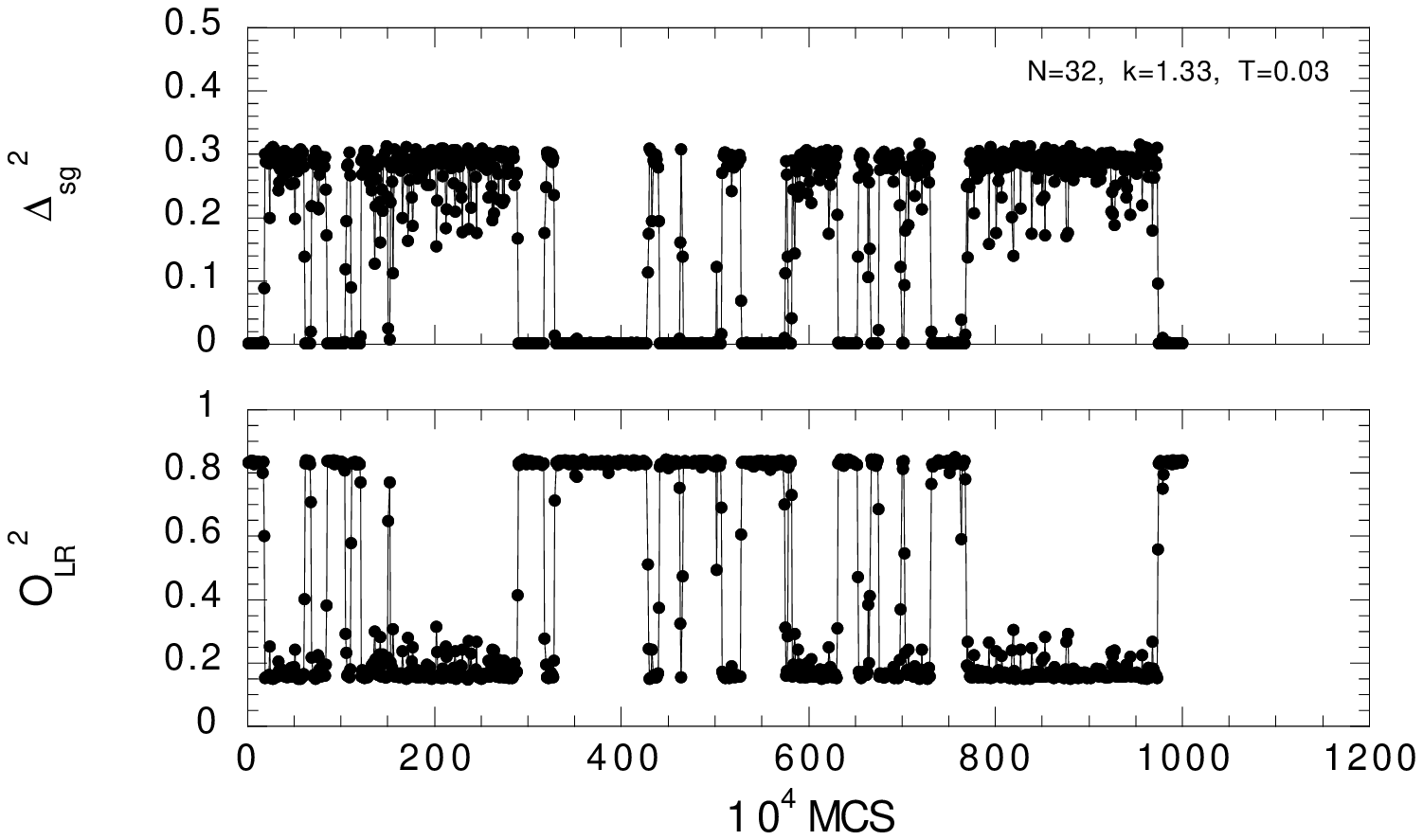}
\vspace*{1mm}
\caption{
The MCS dependence of the bond staggered order $\Delta_{\rm sg}^2$
and the string order $O_{\rm LR}^2$ at $T=0.03$ for $k=1.33$ and $N=32$.
The whole simulation is divided in blocks of $10^4$ MCS
and the thermal averaged value is calculated in each bin.
}
\label{fig:bsg2str2_k1.33}
\end{figure}

\begin{figure}
\centering \epsfxsize=84mm \epsfbox{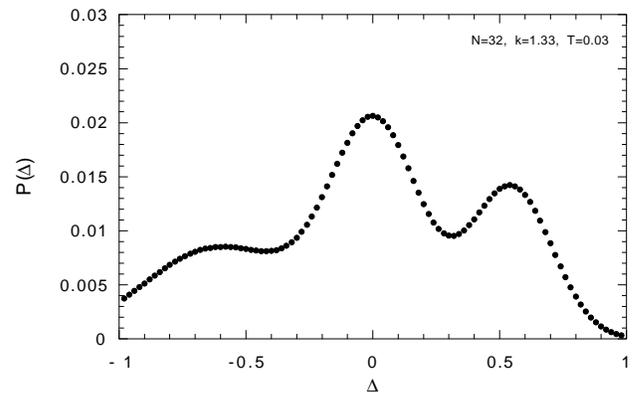}
\vspace*{1mm}
\caption{
The distribution function of the bond distortion
at $T=0.03$ for $k=1.33$ and $N=32$.
}
\label{fig:p-d_k1.33}
\end{figure}

On the other hand,
when the temperature exceeds the energy barrier,
a change of states between the uniform state and the dimerized state occurs
due to the thermal fluctuation.
In Fig.~\ref{fig:bsg2str2_k1.3},
we show the time evolution of the bond staggered order $\Delta_{\rm sg}^2$
and the string order $O_{\rm LR}^2$ for $k=1.3$
at a modestly low temperature $T=0.03$.
Although the coexistence of the two states
is expected in the ground state at $k=1.3$,
the system stays in the dimerized state
through almost the whole simulation.
Namely,
the dimerized state is favored at a finite temperature.
This must be due to an entropy effect.
In order to make it clear,
we study the case of $k=1.33$.
As shown in Fig.~\ref{fig:bsg2str2_k1.33},
we find jumps between the two stable states.
Namely,
the bond configuration takes
a uniform configuration and an alternating arrangement
by taking turns in the simulation,
and correspondingly the spin state takes the Haldane state and the dimer state.
At this value of $k$,
the ground-state energy in the dimerized state is larger than
that in the uniform Haldane state.
However,
the system jumps between the two states almost equally.
This indicates the same free energy of the two states,
which is brought by
the exquisite balance between the quantum fluctuation in the ground state
and the thermal fluctuation at finite temperature.
We indeed find a distribution with three peaks
as shown in Fig.~\ref{fig:p-d_k1.33}.
%
%
%
%

This entropy effect can be understood
from the shape of the effective potential
$E_{\rm g}(\delta)$ shown in Fig.~\ref{fig:e-d-vk}.
If we approximate the potential shape around the bottom of the minima,
we have effective stiffness constants $a_{H}$ and $a_{D}$:
\begin{eqnarray}
E(\delta) =
\left\{
\begin{array}{ll}
\frac{1}{2}a_{H}\delta^2            & (\delta \sim 0)  \\
\frac{1}{2}a_{D}(\delta-\delta_0)^2
               + \Delta E(\delta_0) & (\delta \sim \delta_0), \\
\end{array}
\right.
\end{eqnarray}
where $\Delta E(\delta_0) = \left[E_{\rm g}(\delta_0)-E_{\rm g}(0)\right]/N$.
From Eq.~(\ref{eq:e-d_spin}),
we find $a_{H}=0.560$ and $a_{D}=0.480$
for the critical value $k_c = 1.30$.
The free energy of the system in the harmonic approximation is given by
\begin{equation}
F_{i} = -k_{B}T \ln Z_{i},
\end{equation}
where
\begin{equation}
Z_{i}
= \int_{-\infty}^{\infty} e^{-\beta E_{i}(\delta)} d\delta
= \sqrt{ \frac{2\pi k_{B}T}{a_{i}} }.
\end{equation}
Thus the difference of the free energy is
\begin{eqnarray}
\Delta F
&\equiv&
F_{H} - F_{D} \nonumber \\
&=&
\frac{1}{2} k_{B}T \left( \ln a_{H} - \ln a_{D} \right)
- \Delta E(\delta_0).
\end{eqnarray}
%
%
%
%

\noindent
If we set the value of $k$ to be equal to the critical value $k_c = 1.30$,
then $\Delta E(\delta_0) = 0$ and
$\Delta F = \frac{1}{2}k_{B}T\ln(a_{H}/a_{D})>0$.
Therefore,
the dimerized state is more stable than the uniform Haldane state
at finite temperature.
In order to compensate for the entropy term,
we can set $E_{D}(\delta_0)>0$.
Indeed,
if we use $k=1.33$,
the difference of the free energy becomes very small at $T=0.03$,
which is consistent with the above observation.

\section{COEXISTENT STATE}
\label{sec:coexistent state}

In this section
we study the domain-wall structure
between the uniform Haldane state and the dimerized state.
As we mentioned in the previous section,
in the periodic chain without the inhomogeneity,
the bond configuration takes
the uniform configuration or the dimerized configuration
at low temperature,
and it is rather difficult to see the change between the two states.
In order to see the domain-wall structure,
we fix different bond configurations at each chain end:
a uniform configuration in the left side and
an alternating configuration in the right side.
The bonds between them fluctuate and take a thermal equilibrium configuration.
In order to
investigate spin and bond configurations in the coexistent state,
the amplitude of the bond alternation of the fixed bonds
is set to
the amplitude of the dimerized state
that corresponds to the minimum in the ground-state energy curve
obtained in Sec.~\ref{sec:dimerized ground state}.

\begin{figure}
\centering \epsfxsize=84mm \epsfbox{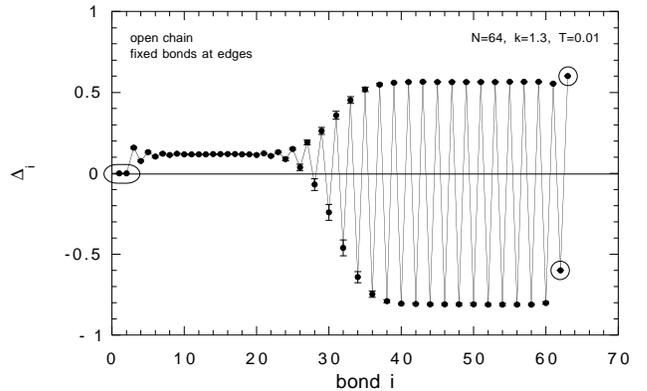}
\vspace*{1mm}
\caption{
The bond configuration in the open chain with fixed boundary bonds
at $T=0.01$ for $k=1.3$ and $N=64$.
The fixed bonds are enclosed in circles.
}
\label{fig:bcf-HD}
\end{figure}

\begin{figure}
\vspace*{-2mm}
\centering \epsfxsize=84mm \epsfbox{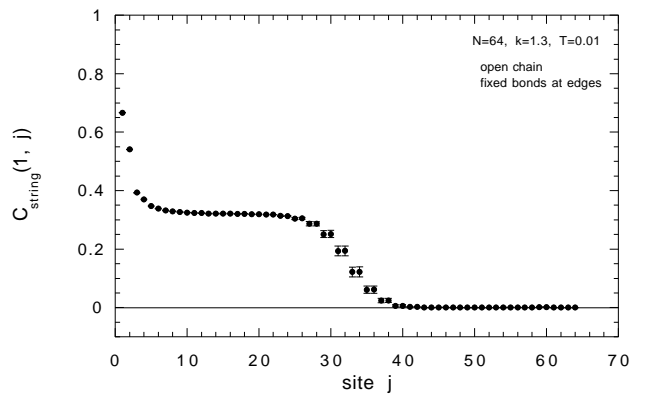}
\vspace*{1mm}
\caption{
The string correlation function
from the left edge site
in the open chain with fixed boundary bonds
at $T=0.01$ for $k=1.3$ and $N=64$.
Fixed boundary bonds are the same as that in Fig.~\ref{fig:bcf-HD}.
}
\label{fig:strcor-HD}
\end{figure}

\begin{figure}
\vspace*{-2mm}
\centering \epsfxsize=84mm \epsfbox{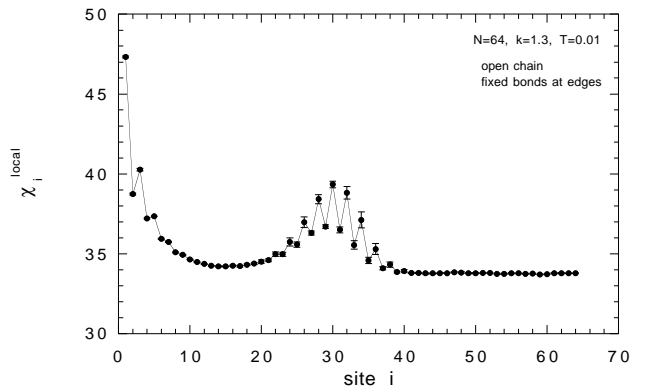}
\vspace*{1mm}
\caption{
The local-field susceptibility $\chi_{i}^{\rm local}$
in the open chain with fixed boundary bonds
at $T=0.01$ for $k=1.3$ and $N=64$.
Fixed boundary bonds are the same as that in Fig.~\ref{fig:bcf-HD}.
}
\label{fig:chi_lf-HD}
\end{figure}

In Fig.~\ref{fig:bcf-HD},
we show the bond configuration at a very low temperature $T=0.01$ for $k=1.3$.
We find
the domain wall between
the uniform and the dimerized configurations.
Namely,
the bonds relax into the uniform configuration in the left side
and into the dimerized configuration in the right side.
In the uniform region,
the lattice shrinks uniformly
where the strength of the exchange coupling is larger than
that of the original uniform chain.
This shift of the exchange coupling causes the magnetic energy gain.
On the other hand,
in the dimerized region,
$\Delta_i$ changes alternately.
However,
the amplitude of negative $\Delta$ increases,
which compensates for the increase in the uniform region.
According to the bond configuration,
the spin state takes a characteristic state in each region.
In Fig.~\ref{fig:strcor-HD},
we show the string correlation function from the left edge site.
In the region of the uniform bond configuration,
the spin state takes the Haldane state and the string correlation persists.
On the other hand,
in the region of the dimerized bond configuration,
the spin state takes the dimer state
and the string correlation decays exponentially.
Thus we find that
the uniform Haldane state and the dimerized state coexist in the chain.
The domain wall between the two states shows a soliton structure.
Finally let us study the local magnetic response of the chain.
In Fig.~\ref{fig:chi_lf-HD},
we show the local-field susceptibility,
\begin{equation}
\left. \chi_{i}^{\rm local}
\equiv \frac{\partial}{\partial h_i} \langle S_i^z\rangle \right |_{h_i=0}
= \int_0^{\beta} d\tau \langle S_i^z(\tau) S_i^z(0) \rangle \, ,
\end{equation}
%
%
%
%
%
%
\noindent
which represents the degree of quantum fluctuation at each site.
We find that
the response is strong at edges of the region of the Haldane state,
which is similar to a pure open Haldane chain~\cite{imp-MiYa}.

\section{SUMMARY}
\label{sec:summary}

In this paper,
we investigated the instability to a dimerized chain
of a one-dimensional $S=1$ antiferromagnetic Heisenberg model
coupled to a lattice distortion
by a quantum Monte Carlo method.
From the $\delta$ dependence of the ground-state energy $E_{\rm g}(\delta)$,
we found that the instability to the dimerized chain
depends on the value of the spin-phonon coupling
unlike the case of $S=\frac{1}{2}$.
We also found the first-order transition
between the uniform Haldane state and the dimerized state.
We investigated spin and bond configurations in the coexistent range.
The system takes turns jumping
between the two states in the simulation.
By studying the temperature dependence of the value of $k$ at the coexistence,
we found that
the dimerized state is favored at finite temperature
due to an entropy effect
that can be explained in a harmonic approximation
of the effective potential function $E(\delta)$.
We also studied the domain-wall structure between the two states
in an open chain fixing the edge bonds to have different states.

As mentioned above,
we found a peculiar character in the case of $S=1$
that is not observed in the case of $S=\frac{1}{2}$.
In a realistic material,
this phenomena could be observed as the spin-Peierls transition
for small values of $k$,
namely, large values of the spin-phonon coupling.
We also expect that
the domain wall between the Haldane and the dimer spin states will be found
in some material.

\acknowledgements

The present study is partially supported by
a Grant-in-Aid from the Ministry of Education, Science,
Sports and Culture of Japan.
The authors are also grateful for the use of the Supercomputer Center
at the Institute for Solid State Physics, University of Tokyo.



\end{document}